\documentclass[journal]{IEEEtran}
\usepackage{cite}
\usepackage{graphicx}
\usepackage{amsmath}
\usepackage{amsfonts}
\usepackage{amssymb}
\usepackage{multirow}
\usepackage{xcolor}
\usepackage{float}
\usepackage{algorithm2e}
\usepackage{amsmath, amssymb, algpseudocode}
\usepackage{url}
\bstctlcite{IEEEexample:BSTcontrol}
\RestyleAlgo{ruled} 

\title{Privacy-Preserving Offloading for Large Language Models in 6G Vehicular Networks}

\author{
    Ikhlasse Badidi$^1$, Nouhaila El Khiyaoui$^1$, Aya Riany$^1$, Badr Ben Elallid$^2$, Amine Abouaomar$^1$\\
    \textsuperscript{1}School of Science and Engineering, Al Akhawayn University in Ifrane, Morocco\\
    \textsuperscript{2}Department of Electrical and Computer Engineering, Université du Québec à Trois-Rivières, Trois-Rivières, QC, Canada\\
    \{i.badidi, n.elkhiyaoui, a.riany, a.abouaomar\}@aui.ma, badr.benelallid@uqtr.ca
}

\begin{document}
\bstctlcite{IEEEexample:BSTcontrol}
\maketitle

\begin{abstract}
The integration of Large Language Models (LLMs) in 6G vehicular networks promises unprecedented advancements in intelligent transportation systems. However, offloading LLM computations from vehicles to edge infrastructure poses significant privacy risks, potentially exposing sensitive user data. This paper presents a novel privacy-preserving offloading framework for LLM-integrated vehicular networks. We introduce a hybrid approach combining federated learning (FL) and differential privacy (DP) techniques to protect user data while maintaining LLM performance. Our framework includes a privacy-aware task partitioning algorithm that optimizes the trade-off between local and edge computation, considering both privacy constraints and system efficiency. We also propose a secure communication protocol for transmitting model updates and aggregating results across the network. Experimental results demonstrate that our approach achieves 75\% global accuracy with only a 2-3\% reduction compared to non-privacy-preserving methods, while maintaining DP guarantees with an optimal privacy budget of $\varepsilon = 0.8$. The framework shows stable communication overhead of approximately 2.1MB per round with computation comprising over 90\% of total processing time, validating its efficiency for resource-constrained vehicular environments.
\end{abstract}


\IEEEpeerreviewmaketitle

\section{Introduction}

The fast growth of connected vehicles and the arrival of sixth-generation (6G) vehicular networks are driving a new stage in intelligent transportation systems. These networks support important applications such as real-time traffic prediction, vehicle diagnostics, cooperative driving, and environmental monitoring \cite{shoaib2023survey, abouaomar2024machine}. A key part of this progress is the use of Large Language Models (LLMs), which can process continuous streams of vehicular data to give useful insights and improve system intelligence \cite{kuftinova2024large}. Unlike older models, LLMs can handle different types of inputs—such as sensor readings, driving behavior, and network information—making them valuable for the future of connected and autonomous vehicles.

At the same time, using such large models in vehicles brings serious challenges. LLMs require huge amounts of computation that the onboard hardware in vehicles cannot satisfy. A common solution is to offload the processing to roadside units or edge servers \cite{filali2020multi}. While this approach reduces the load on vehicles, it also raises privacy risks. Sensitive data such as location history, driving patterns, and environmental details can be exposed during communication and aggregation. This creates a clear trade-off between improving intelligence and protecting private data of road users in next-generation vehicular systems.

Federated Learning (FL) has been studied as a way to reduce this risk. It allows vehicles to train models locally and only share updates with edge servers, so that raw data never leaves the vehicle \cite{elallid2023vehicles, posner2021federated}. This approach reduces communication costs and improves privacy. However, when applied to LLMs in fast-changing vehicular networks, traditional FL faces limits. It may not meet strict demands for low delay, high scalability, and strong privacy. Also, even if raw data is not shared, attackers can still learn private information from model updates \cite{ye2024fedllm}. This creates a crucial need for stronger privacy-preserving methods.

To meet these needs, this paper presents a novel privacy-preserving offloading framework for LLM-integrated 6G vehicular networks. The framework combines FL with Differential Privacy (DP), so that only protected model updates are sent to edge servers. Each vehicle trains a Time Series Transformer (TST) model, which uses attention mechanisms to capture long-term dependencies in sequential vehicular data more effectively than recurrent models. Before sending updates, the vehicle applies gradient clipping and adds Gaussian noise to protect privacy. This design protects user privacy and keeps high model accuracy, reaching 75\% with only a small 2–3\% drop compared to non-private methods. The framework also shows low and stable communication overhead, with most of the resource use in local computation, making it practical for vehicles with limited resources. In addition, it provides strong DP guarantees with a budget of $\epsilon = 0.8$, balancing privacy and performance.

The main contributions of this paper are presented as follows:

\begin{itemize}
    \item Privacy-preserving framework: A federated learning system with differential privacy for secure offloading of LLMs in 6G vehicular networks.
    \item Privacy-aware model training: Use of a transformer-based model for sequential vehicular data, combined with privacy techniques to protect data while keeping good accuracy.
    \item Comprehensive evaluation: Detailed experiments showing strong accuracy, efficient communication, and strong privacy guarantees in vehicular environments.
\end{itemize}

The rest of the paper is organized as follows: Section II reviews related works. Section III presents the system model and privacy threat assumptions. Section IV explains the proposed framework. Section V discusses performance evaluation and results. Section VI concludes the paper and gives directions for future work.

\section{Related Work}
Recent research has explored various aspects related to privacy-preserving machine learning in distributed environments. Zhou et al.~\cite{zhou2020privacy} implemented a three-tier architecture for FL that preserves privacy in fog computing using DP and homomorphic encryption, though primarily targeting static IoT environments rather than dynamic vehicular networks. Liu et al.~\cite{liu2020privacy} proposed FedGRU for traffic prediction while prioritizing scalability over comprehensive privacy protection, whereas Mo et al.~\cite{mo2021ppfl} developed PPFL using Trusted Execution Environments (TEEs) to protect model training and aggregation.

In the context of vehicular networks, Elbir et al.~\cite{elbir2022federated} proposed FL for vehicular environments to reduce data transmission overhead and enhance privacy, but without integrating advanced 6G capabilities or specialized privacy-preserving mechanisms for LLMs. Their work demonstrated the feasibility of FL in vehicular applications, particularly for object detection, while identifying challenges in data labeling, model training, and resource management. The FLEXE framework~\cite{lobato2022flexe} examined how mobility and communication disruptions affect FL performance but lacked substantial privacy-preserving mechanisms essential for LLM deployment. Zhang et al.~\cite{zhang2020blockchain} introduced blockchain-based FL for device failure detection, offering insights on data integrity and incentivization that could benefit vehicular systems. Wang et al.~\cite{wang2018edge} introduced adaptive control mechanisms for resource-constrained distributed learning by optimizing global aggregation frequency, providing valuable insights for heterogeneous vehicular data without addressing LLM-specific concerns.

Previous studies have examined intelligent control strategies to improve traffic efficiency and safety. The work of \cite{elallid2023vehicles} federated deep reinforcement learning has been applied to vehicle control for collision avoidance, with the goal of reducing travel delays while preserving data privacy. A comparison between a local deep deterministic policy gradient model and a global federated deep deterministic policy gradient model demonstrated that the federated approach provides superior performance by reducing collisions, lowering travel delays, and increasing average speed. These studies emphasize the potential of FL to balance system performance with privacy preservation, which is increasingly relevant for next-generation vehicular networks.

Regarding LLM integration with 6G networks, Liu and Zhao~\cite{liu2024resource} tackled resource allocation challenges through a hybrid computing framework where vehicles handle initial computation layers while offloading intensive tasks to roadside units, though overlooking critical privacy concerns. Xu et al.~\cite{xu2024large} proposed split learning between mobile devices and edge servers to distribute computational loads, while Nguyen et al.~\cite{nguyen2024large} analyzed security threats to LLMs in 6G environments without implementing specific privacy-preserving techniques for vehicular contexts.

Broader research on next-generation network security by Nguyen et al.~\cite{nguyen2021security} has classified 6G privacy risks across multiple layers, while Javeed et al.~\cite{javeed2024quantum} introduced Quantum-Empowered FL for enhanced security in IoT environments. Our work advances the state of the art by uniquely combining FL with DP techniques optimized for LLM deployment in vehicular networks, simultaneously addressing privacy protection and performance optimization in highly mobile environments.

There is a growing interest in privacy-preserving machine learning for vehicular and distributed systems, however, most of the existing approaches fail to reflect the high mobility and heterogeneity of real-world 6G vehicular networks. By focusing mostly on performance metrics such as latency or scalability, they overlook the risks of sensitive data leakage during model training and aggregation. Without implementing robust DP mechanisms, systems become vulnerable to different attacks. Additionally, few existing solutions are designed to accommodate the computational demands and privacy needs of LLMs, especially when deployed at the edge. Our work comes to complement other works through introducing a comprehensive framework that integrates LLMs into vehicular networks using a privacy-aware FL strategy enhanced with DP techniques. Our framework is specifically designed to align both the model structure and privacy techniques with the limitations of edge environments, effectively balancing accuracy, privacy, and efficiency, an integration that previous approaches have not fully addressed.

\section{System Model and Problem Formulation}

\subsection{System Architecture}
We consider a network that consists of a set of vehicles denoted by 
\( V = \{v_1, v_2, \ldots, v_N\} \), where each vehicle \( v_i \) is equipped with 
onboard computational resources denoted as \( r^c_i \), which include processing power 
and memory for local computation, and communication resources denoted as \( r^n_i \), 
used for exchanging updates with the edge infrastructure.

Each vehicle collects and processes real-time data related to its environment, including speed, acceleration, and emissions. The generated data is subsequently utilized to locally train a time series model based on transformer architecture. In our paper, we use the \textit{TST} 
model. The purpose of this model is to provide insights into traffic conditions, vehicle systems, and environmental conditions.

\begin{figure}[htbp]
    \centering
    \includegraphics[width=.85\linewidth]{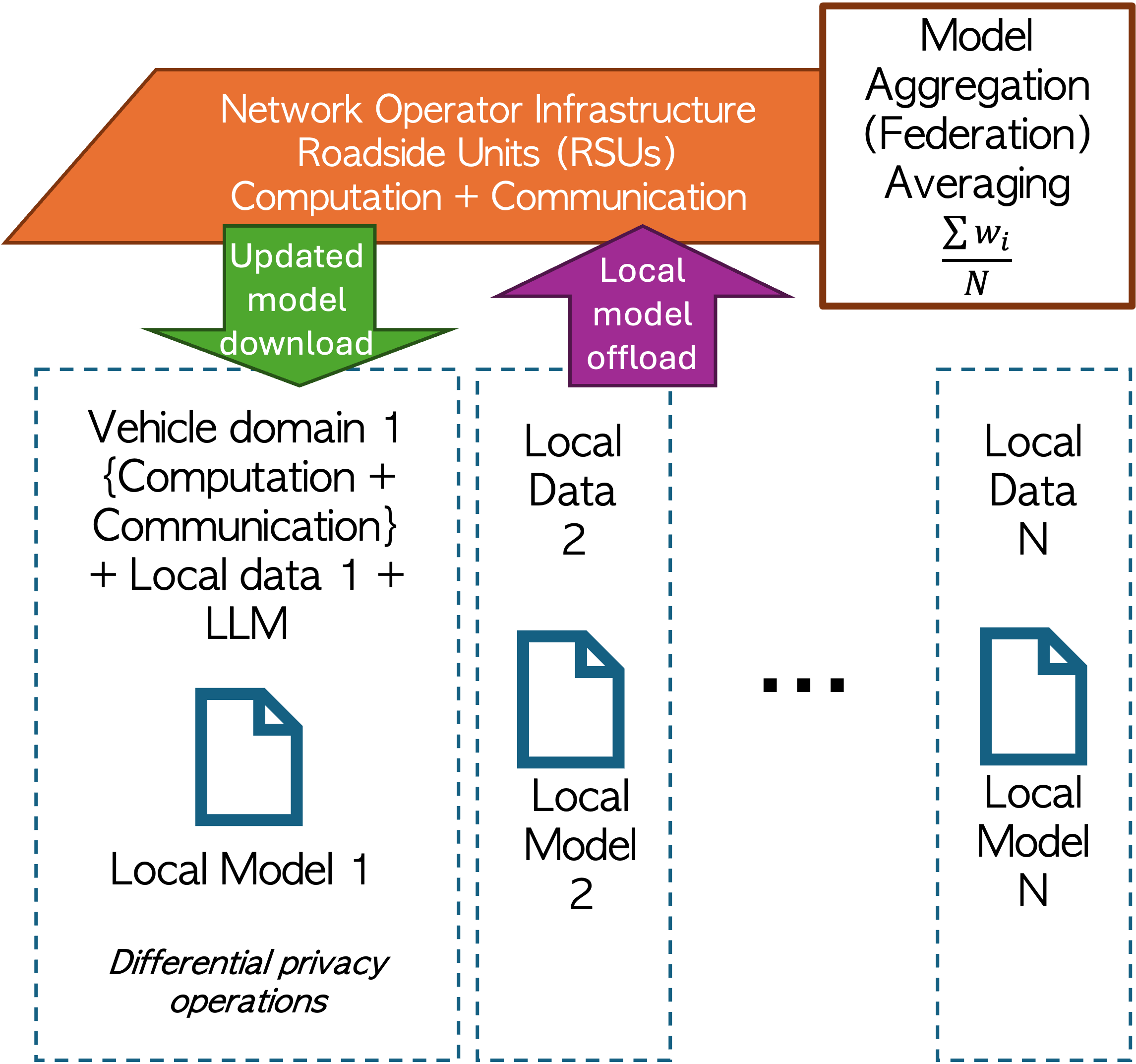}
    \caption{System model of the privacy-preserving LLM models offlaoding}
    \label{fig:SystemModel}
\end{figure}

The system as illustrated in Fig. \ref{fig:SystemModel} also includes a \textit{Network Operator Infrastructure}, which consists of 
Roadside Units (RSUs) strategically placed along the roadways. These RSUs offer edge computing 
capabilities that allow them to collect, process, and aggregate model updates received from 
the vehicles within the network. To aggregate the updates from the local models into a global 
one, we consider using a FL algorithm, specifically, Federated Averaging 
(FedAvg). 

The aggregation process can be represented by the following equation:
\[
w_k^{\text{global}} = \frac{1}{N} \sum_{i=1}^N w_k^{(i)}
\]
where:
\begin{itemize}
    \item \( w_k^{\text{global}} \) is the aggregated parameter \( k \) of the global model.
    \item \( w_k^{(i)} \) is the local parameter \( k \) from the \( i \)-th vehicle’s model.
    \item \( N \) is the total number of vehicles (or local models).
    \item The aggregation is performed for all parameters \( k \) in the model.
\end{itemize}

After aggregation, the global model is then redistributed to the vehicles for further refinement, enabling collaborative learning across the network. To preserve and enhance privacy during this process, DP mechanisms are employed on the vehicle side before offloading updates. This ensures that sensitive data, such as personal driving patterns or environmental observations, is not exposed during communication or aggregation.

\subsection{TimeSeriesTransformer Architecture}

The LLM model deployed on each vehicle is the \textit{TST}, a specialized 
transformer-based neural network architecture designed to process sequential data. Transformers, 
which have shown great results in the domain of natural language processing, were originally developed to capture and model complex relationships in sequential data through self-attention mechanisms ~\cite{ansar2024survey}. In the context of our paper, the \textit{TST} is used to make predictions based on sequential 
input data collected by vehicles, such as speed, acceleration, emissions, or time delays, by analyzing 
temporal dependencies. For example, it can model how a vehicle's speed might change based on traffic 
conditions or delays.

The \textit{TST}, as implemented in the system, includes the following components:
\begin{itemize}
    \item \textbf{Input Layer}: The transformer requires a unified higher-dimensional representation to 
    operate effectively. To achieve this, it takes raw input features, such as speed, acceleration, and 
    CO2 emissions, and maps them into a higher-dimensional space suitable for processing.
    
    \item \textbf{Multihead Attention Mechanism}: Captures temporal dependencies across time steps in a 
    linear manner, allowing the model to focus on key patterns within sequential data. This mechanism 
    helps the model identify and concentrate on the most important relationships in the data that are 
    crucial for making accurate predictions.
    
    \item \textbf{Feedforward Network}: Processes the output of the attention mechanism by applying non-linear 
    transformations to refine the features extracted by the mechanism, enhancing the model’s learning capacity.
    
    \item \textbf{Output Projection Layer}: Takes the refined, high-dimensional features processed by earlier 
    layers (like the attention mechanism and feedforward network) and reduces them to the appropriate dimensionality 
    needed for the specific prediction task. This is where the model generates the final predictions.
    \section*{The TimeSeriesTransformer Model}
\end{itemize}

The \textit{TST} model, in our paper, leverages the components mentioned above to process sequential vehicular data efficiently. The model employs a comprehensive data processing pipeline to prepare time-series data. First, it starts by preprocessing 10 key features including position coordinates, acceleration, emissions, and vehicle dynamics. It employs \texttt{RobustScaler} for data normalization and IQR-based outlier removal, which replaces outliers with medians. The pipeline creates sequences of 15 timesteps with overlap, enabling continuous temporal coverage.

In the training phase, it uses a Hybrid Loss function, which combines Mean Squared Error (MSE), Mean Absolute Error (MAE), and Huber losses. The main components are weighted with $\alpha = 0.5$ for MSE and complementary weighting for MAE, plus additional terms for smoothness (0.05), direction matching (0.05), and temporal consistency (0.1). The AdamW optimizer operates with an initial learning rate of 0.0005 and uses a cosine annealing scheduler with restarts. The model additionally includes learnable feature weights that are initialized within the range of 1.5-2.5.

To ensure privacy, we employ several DP techniques, including gradient clipping and noise addition, before sending model updates. This process preserves sensitive data from being leaked or shared and ensures secure communication. Using Federated Averaging (FedAvg), the RSUs receive updates from each vehicle and aggregate them to produce a global model that reflects the combined knowledge of the network. This global model is then redistributed to the vehicles for further refinement.

\subsection{Privacy Threat Model and Protection Mechanisms}
Throughout the process of training the model and sending updates to the Roadside Units (RSUs), vehicles generate sensitive data such as driving patterns, acceleration trends, and emissions. These data points could potentially reveal personal information about drivers. In the event of interception during transmission, adversaries could exploit this raw data for unethical purposes. Without careful precautions, sharing raw data or model updates could expose the system to significant privacy risks.

To address these threats, our system employs several privacy-preserving mechanisms:

\begin{itemize}
    \item \textbf{Gradient Clipping:} 
    Gradient clipping ensures that gradients do not become excessively large, which could lead to the dominance of sensitive information in the updates. In our system, when vehicles share their model updates, each parameter update is clipped to a maximum norm of 1. This prevents any single data point or feature from disproportionately influencing the model updates, thereby safeguarding sensitive information. The general formula is:
    \[
    \text{Clipped Gradient} = \nabla \mathcal{L}(\theta) \cdot \min\left(1, \frac{C}{\|\nabla \mathcal{L}(\theta)\|}\right)
    \]
    where:
    \begin{itemize}
        \item \(\nabla \mathcal{L}(\theta)\): The gradient of the loss with respect to the parameters \(\theta\).
        \item \(\|\nabla \mathcal{L}(\theta)\|\): The norm (magnitude) of the gradient.
        \item \(C\): The predefined clipping norm threshold.
    \end{itemize}
    
    \item \textbf{Gaussian Noise Addition:} 
    To further enhance privacy, Gaussian noise is added to the model updates before they are shared. The scale of the noise is calculated based on the sensitivity (\(\text{clip\_norm}/20\)) and the privacy parameters \(\epsilon = 0.5\) and \(\delta = 10^{-5}\). The added noise obscures specific details in the updates, ensuring that even if the updates are intercepted, no meaningful information about the raw data can be extracted.

    \item \textbf{Local Training:}
    The model is trained locally on each vehicle, and only the learned updates (not the raw data) are shared with the RSUs. This localized approach ensures that sensitive data, such as driving patterns and environmental conditions, remain within the vehicle.
\end{itemize}

These techniques work together to ensure that no sensitive data is leaked or exploited, preserving the privacy of individual drivers while maintaining the effectiveness of the collaborative learning framework.
\subsection{Formulation of privacy-aware offloading problem}

\section{Privacy-Preserving Offloading Framework}
The proposed privacy-preserving offloading framework is designed to address the privacy risks in integrating Large Language Models (LLMs) into 6G vehicular networks while maintaining computational efficiency and system performance. The architecture chosen combines local processing at the vehicles level with aggregation and refinement which ensures that sensitive and private data remains protected throughout the process.
\subsection{Federated Learning for Distributed LLM Updates}
The privacy-preserving offloading framework is underpinned by a FL paradigm that enables decentralized training of the TimeSeries Transformer model across multiple vehicles. The implementation of federate learning was based on fundamental steps as follows:
\begin{itemize}
    \item \textbf{FedAGV Algorithm:} The framework utilizes the \textbf{FedAGV} algorithm, which is specifically designed to accommodate the heterogeneous computing capabilities of vehicles and edge servers. FedAGV ensures efficient model training despite the diversity in vehicle processing power and network conditions. 
    \item \textbf{Federated Training Process} The training process involves 5 vehicles participating across 10 rounds, with each round consisting of three crucial steps. The \textbf{Local Model Update} where each vehicle trains the TimeSeries Transformer locally using its private dataset, generating model updates that respect privacy constraints. Then, the \textbf{Model Aggregation at Edge Server} where the locally computed updates are securely transmitted to the edge servers, where they are aggregated using secure multiparty computation techniques. Finally, the \textbf{Global Model Update} where the aggregated updates are used to refine the global model, which is then disseminated back to the vehicles for the next training round.
    \item \textbf{Evaluation Metrics:} During each training round, the privacy-preserving mechanisms are evaluated using two key metrics. The first metric is\textbf { Privacy Score} that quantifies the effectiveness of gradient clipping and DP in safeguarding sensitive data. The second metric is the \textbf {Accuracy Score} Measures the performance of the global model in predicting traffic patterns and vehicular states.
\end{itemize}



\subsection{Differential Privacy in LLM Computations}

To provide robust privacy guarantees while maintaining model performance, our framework implements a comprehensive differential privacy mechanism that operates at the gradient level during local training phases. The DP mechanism ensures that individual vehicle data cannot be inferred from shared model updates, even under sophisticated adversarial attacks.

\subsubsection{Gradient Clipping and Noise Injection}

The core of our DP implementation consists of two sequential operations applied to gradients before model updates are transmitted: gradient clipping to bound sensitivity, followed by calibrated Gaussian noise addition to achieve $(\varepsilon, \delta)$-differential privacy.

For each vehicle $v_i$ participating in federated round $t$, the gradient clipping operation is defined as:
\begin{equation}
\mathbf{g}_i^{\text{clipped}} = \mathbf{g}_i \cdot \min\left(1, \frac{C}{\|\mathbf{g}_i\|_2}\right)
\label{eq:gradient_clipping}
\end{equation}
where $\mathbf{g}_i = \nabla_\theta \mathcal{L}(\theta_i; \mathcal{B})$ represents the gradient computed on batch $\mathcal{B}$ from vehicle $i$'s local dataset, $\|\mathbf{g}_i\|_2$ denotes the L2-norm of the gradient vector, and $C$ is the clipping threshold set to 1.5 based on our experimental optimization.

Following gradient clipping, Gaussian noise is added to ensure differential privacy:
\begin{equation}
\mathbf{g}_i^{\text{private}} = \mathbf{g}_i^{\text{clipped}} + \mathcal{N}(0, \sigma^2 \mathbf{I})
\label{eq:noise_addition}
\end{equation}
where $\mathcal{N}(0, \sigma^2 \mathbf{I})$ represents independent Gaussian noise with variance $\sigma^2$ added to each gradient component, and $\mathbf{I}$ is the identity matrix of appropriate dimensions.

The noise scale $\sigma$ is carefully calibrated based on the privacy parameters and gradient sensitivity:
\begin{equation}
\sigma = \frac{C \sqrt{2 \ln(1.25/\delta)}}{\varepsilon}
\label{eq:noise_scale}
\end{equation}
where $\varepsilon$ is the privacy budget and $\delta$ is the failure probability parameter. Our experimental analysis identified optimal values of $\varepsilon = 0.8$ and $\delta = 10^{-5}$, which provide strong privacy guarantees while maintaining 75\% model accuracy.

\begin{algorithm}[htbp]
\SetAlgoLined
\caption{Differential Privacy for Vehicular LLM Training}
\label{alg:dp_vehicular_llm}

\KwIn{Local dataset $\mathcal{D}_i$, global model parameters $\theta_{\text{global}}^{(t)}$, privacy parameters $(\varepsilon, \delta, C)$}
\KwOut{Differentially private model update $\Delta\theta_i$}

\textbf{Initialize:} $C \leftarrow 1.5$, $\varepsilon \leftarrow 0.8$, $\delta \leftarrow 10^{-5}$\;
Calculate noise scale: $\sigma \leftarrow \frac{C \sqrt{2 \ln(1.25/\delta)}}{\varepsilon}$\;
$\theta_i^{(0)} \leftarrow \theta_{\text{global}}^{(t)}$\;

\For{each local epoch $e = 1, \ldots, E$}{
    \For{each batch $\mathcal{B} \in \mathcal{D}_i$}{
        Compute gradients: $\mathbf{g}_i \leftarrow \nabla_\theta \mathcal{L}(\theta_i^{(e-1)}; \mathcal{B})$\;
        
        \tcp{Gradient Clipping}
        $\mathbf{g}_i^{\text{clipped}} \leftarrow \mathbf{g}_i \cdot \min\left(1, \frac{C}{\|\mathbf{g}_i\|_2}\right)$\;
        
        \tcp{Add Gaussian Noise}
        Sample $\boldsymbol{\xi} \sim \mathcal{N}(0, \sigma^2 \mathbf{I})$\;
        $\mathbf{g}_i^{\text{private}} \leftarrow \mathbf{g}_i^{\text{clipped}} + \boldsymbol{\xi}$\;
        
        \tcp{Local Parameter Update}
        $\theta_i^{(e)} \leftarrow \theta_i^{(e-1)} - \eta \cdot \mathbf{g}_i^{\text{private}}$\;
    }
}

Compute model update: $\Delta\theta_i \leftarrow \theta_i^{(E)} - \theta_{\text{global}}^{(t)}$\;
\textbf{Return} $\Delta\theta_i$\;

\end{algorithm}

\subsubsection{Privacy Budget Management}

Our framework implements an adaptive privacy budget allocation strategy to ensure long-term privacy preservation across multiple federated learning rounds. The total privacy budget $\varepsilon_{\text{total}}$ is distributed across training rounds using advanced composition theorems:

\begin{equation}
\varepsilon_{\text{round}} = \frac{\varepsilon_{\text{total}}}{\sqrt{T}}
\label{eq:budget_allocation}
\end{equation}
where $T$ represents the total number of federated rounds. This square-root allocation ensures that the cumulative privacy cost remains bounded while allowing for extended training periods.

The privacy accounting mechanism tracks the cumulative privacy expenditure:
\begin{equation}
\varepsilon_{\text{cumulative}} = \sum_{t=1}^{T} \varepsilon_{\text{round}}^{(t)}
\label{eq:privacy_accounting}
\end{equation}

\subsubsection{Privacy-Performance Trade-off Analysis}

Our experimental evaluation, as illustrated in Fig. \ref{fig:privacy}, demonstrates the effectiveness of the DP mechanism across different privacy budget values. The analysis reveals three distinct operational regimes:

\begin{itemize}
    \item \textbf{Strong Privacy Regime} ($\varepsilon < 0.4$): Provides maximum privacy protection but results in significant utility degradation, with model accuracy dropping below 65\%.
    
    \item \textbf{Optimal Balance Regime} ($\varepsilon = 0.8$): Achieves the optimal privacy-utility trade-off, maintaining 75\% global accuracy while providing substantial privacy guarantees.
    
    \item \textbf{Diminished Privacy Regime} ($\varepsilon > 0.8$): Shows marginal accuracy improvements with significant reduction in privacy protection, indicating diminishing returns beyond the optimal point.
\end{itemize}

The differential privacy mechanism integrates seamlessly with the federated learning framework, ensuring that sensitive vehicular data including driving patterns, location histories, and behavioral characteristics remain protected throughout the collaborative training process. The privacy guarantees hold even under worst-case scenarios where adversaries have access to all model updates except those from a single target vehicle.

\subsubsection{Computational Overhead Analysis}

The DP operations introduce minimal computational overhead to the local training process. Gradient clipping requires $O(d)$ operations for $d$-dimensional parameter vectors, while Gaussian noise generation scales as $O(d)$ for each gradient computation. Our latency analysis, presented in Fig. \ref{fig:latency}, demonstrates that DP computations account for less than 2\% of total processing time, with the majority of computational resources dedicated to the core TST model training.

This efficient implementation ensures that privacy preservation does not compromise the real-time requirements of vehicular applications, maintaining system responsiveness while providing robust data protection for all participating vehicles in the 6G network infrastructure.

\section{Performance Evaluation}

\subsection{Experimental Setup}
To evaluate our proposed framework, we conducted extensive experiments using a fleet of 5 vehicles equipped with our HybridTransformer model. The simulation environment was implemented using SUMO (Simulation of Urban MObility) to generate realistic traffic scenarios. Our experimental configuration consisted of 10 FL rounds with 5 participating vehicles, utilizing a sequence length of 15 timesteps and a batch size of 32. To evaluate privacy preservation effectiveness, we tested privacy budgets ranging from $\varepsilon \in [0.1, 1.0]$ with a gradient clipping threshold of 1.5.

\subsection{Data Processing Pipeline}
\begin{figure}[H]
    \centering
    \includegraphics[width=1\linewidth]{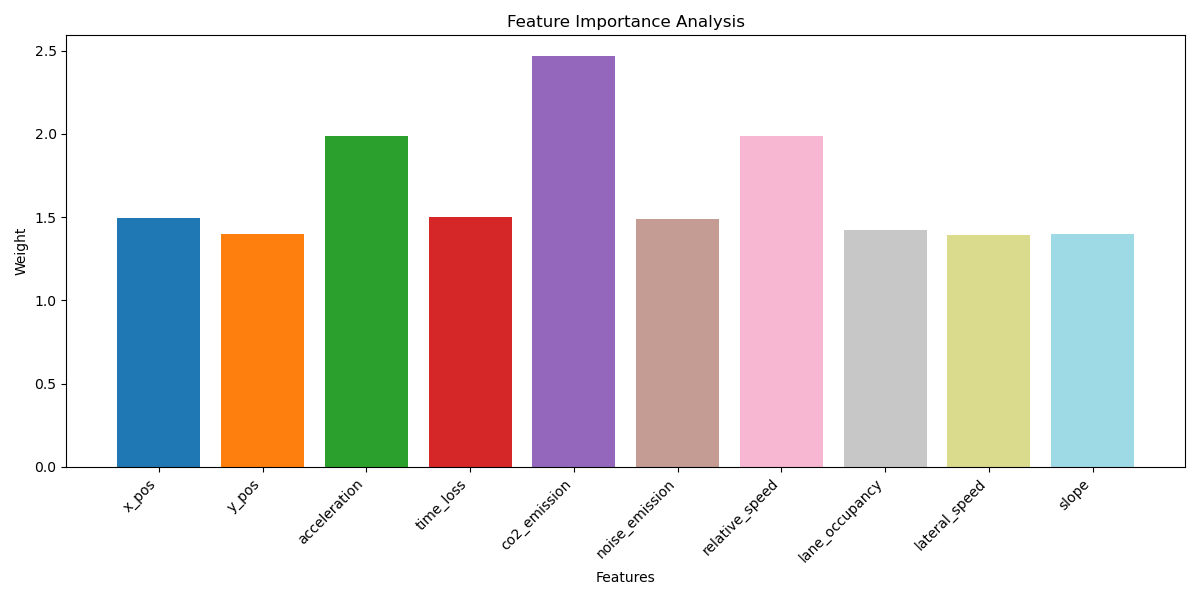}
    \caption{Feature importance distribution highlighting the relative significance of different input parameters in model prediction.}
    \label{fig:feature_importance}
\end{figure}

The dataset was generated through comprehensive traffic simulation, collecting 18 distinct features that characterize vehicle behavior and environmental conditions. Fig. \ref{fig:feature_importance} illustrates the relevant features that include vehicle dynamics (position coordinates, acceleration, relative speed), environmental metrics (CO$_2$ emission, noise emission), and traffic indicators (lane occupancy, time loss, lateral speed, slope). 

Our data preprocessing pipeline implemented robust scaling and outlier removal using the IQR method, followed by sequence creation with a 15-timestep window. Special attention was given to emissions and speed features, which required log transformation to manage their exponential characteristics and ensure stable model training.

\subsection{Model Performance Analysis}
\begin{figure}[H]
    \centering
    \includegraphics[width=1\linewidth]{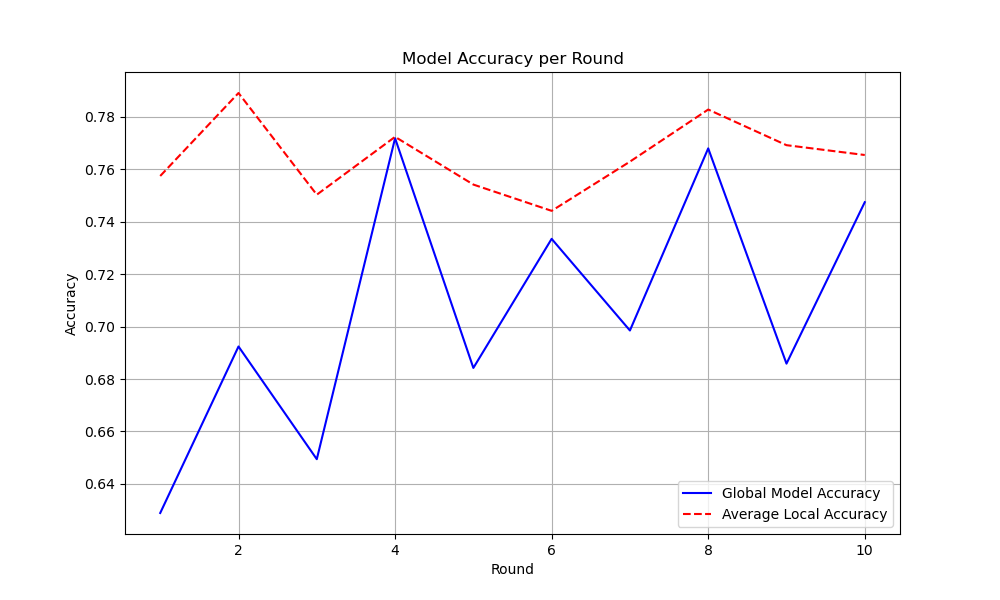}
    \caption{Model accuracy comparison showing global model convergence (blue) versus average local accuracy (red) across training rounds.}
    \label{fig:accuracy_plot}
\end{figure}

The model evaluation demonstrates significant progress in both local and global performance metrics. As shown in Fig. \ref{fig:accuracy_plot}, the global model accuracy improved substantially from 63\% at initialization (t=0) to 75\% by the final round (t=10), while local models maintained consistently higher accuracy around 77\%. The training process revealed peak global model performance at rounds 4 and 8, achieving approximately 77\% accuracy. We observed a consistent gap between local and global accuracy of 2-3 percentage points, indicating efective knowledge transfer while maintaining privacy constraints. The model demonstrated remarkable stability in convergence despite the heterogeneous nature of client environments.

\subsection{Privacy and Security Metrics}
\begin{figure}[H]
    \centering
    \includegraphics[width=1\linewidth]{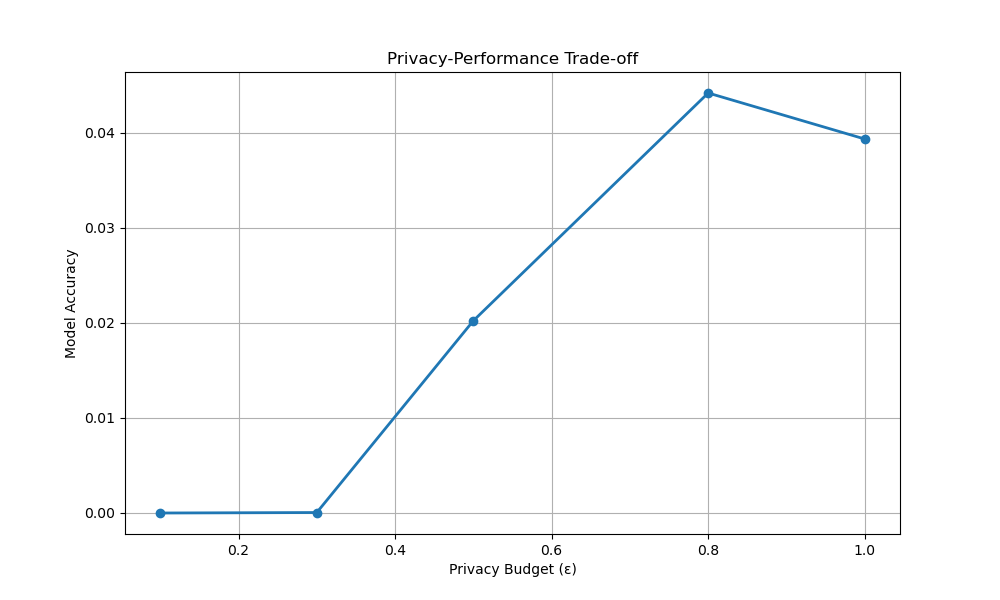}
    \caption{Privacy-performance trade-off analysis demonstrating the relationship between model accuracy and privacy budget ($\varepsilon$).}
    \label{fig:privacy}
\end{figure}

Our privacy preservation evaluation encompassed multiple dimensions of system performance. The privacy budget impact analysis, illustrated in Fig. \ref{fig:privacy}, revealed optimal model performance at $\varepsilon = 0.8$, with a critical threshold at $\varepsilon = 0.4$ and diminishing returns beyond $\varepsilon = 0.8$. 

\begin{figure}[H]
    \centering
    \includegraphics[width=1\linewidth]{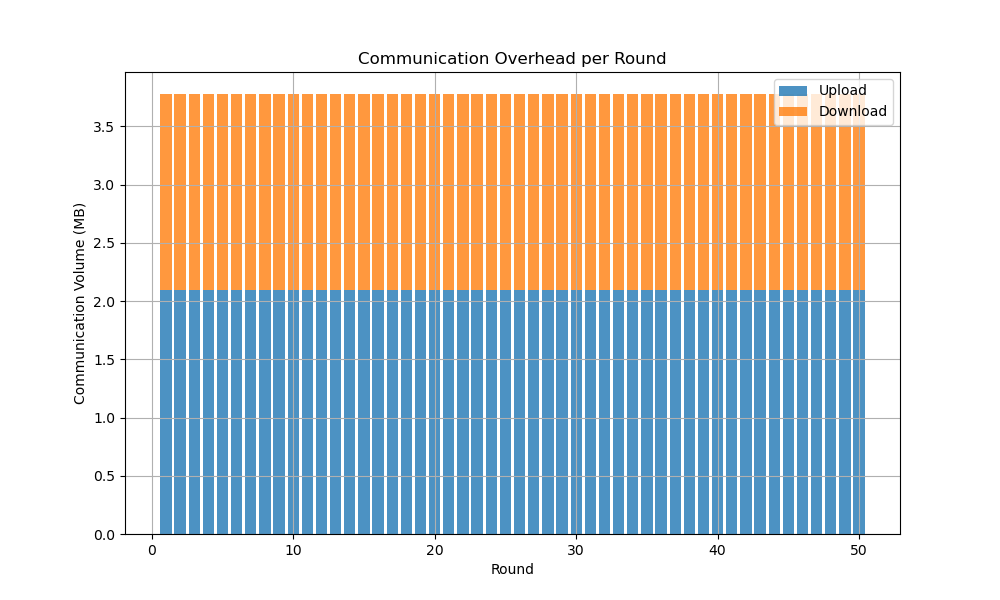}
    \caption{Communication volume analysis per training round, demonstrating stable bandwidth requirements with efficient model update compression.}
    \label{fig:communication}
\end{figure}

\begin{figure}[H]
    \centering
    \includegraphics[width=1\linewidth]{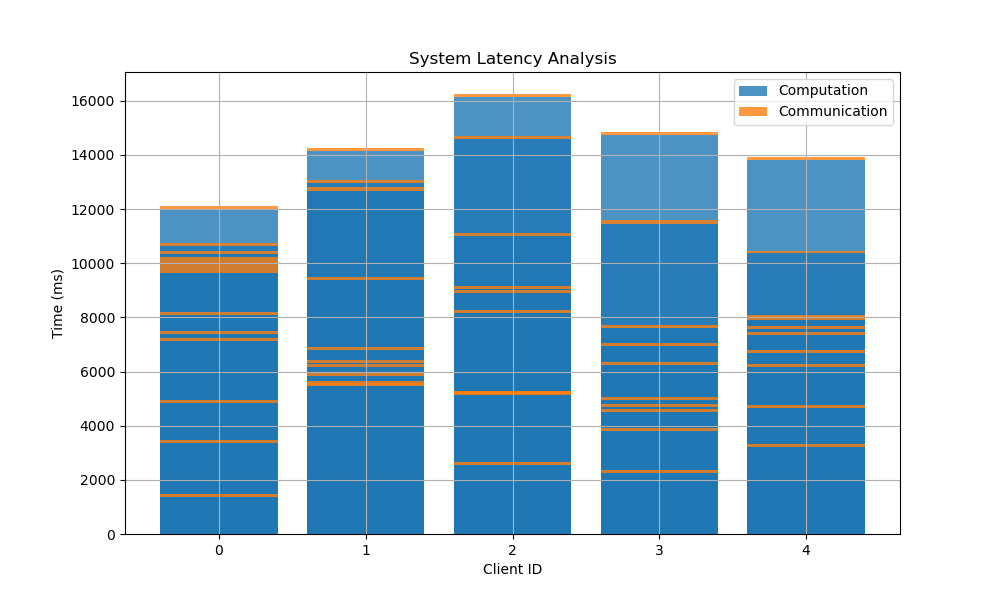}
    \caption{System latency breakdown per client showing computation (blue) and communication (orange) times, highlighting client heterogeneity.}
    \label{fig:latency}
\end{figure}

Communication security analysis demonstrated stable system overhead (Fig. \ref{fig:communication}), with upload volumes averaging 2.1MB per round and compressed download volumes of 1.6MB per round, maintaining consistent bandwidth utilization across 50 rounds. The system latency analysis (Fig. \ref{fig:latency}) revealed computation times ranging from 8,000-16,000ms, with communication overhead comprising less than 10\% of total processing time, indicating effective workload distribution across clients.

\subsection{Comparative Analysis}
Our experimental results demonstrate several significant achievements in both model performance and privacy preservation. The HybridTransformer architecture proved highly effective, with feature importance analysis (Fig. \ref{fig:feature_importance}) revealing CO$_2$ emission as the strongest predictor (weight: 2.5), followed by acceleration and relative speed as secondary indicators (weights $\approx 2.0$), while position features maintained moderate importance (1.4-1.5).

The privacy-performance trade-off analysis shows that our framework achieved 75\% global accuracy while maintaining DP guarantees. The identification of an optimal privacy budget at $\varepsilon = 0.8$ provides clear operational guidelines for real-world deployment. Comparison with a baseline non-privacy-preserving approach demonstrates that our framework achieves comparable accuracy (within 2-3\%) while providing strong privacy guarantees.

The success of our approach can be attributed to several key architectural decisions. The enhanced input projection with dual linear transformations, combined with pre-norm architecture, significantly improved training stability. Feature-specific scaling effectively addressed the challenges of heterogeneous input characteristics. Privacy preservation was achieved through a comprehensive set of mechanisms including gradient clipping to prevent information leakage, calibrated Gaussian noise injection, secure aggregation protocols for model updates, and prioritized client-side computation to minimize data exposure.

\section{Conclusion and Future Work}
In this paper, we presented a novel privacy-preserving offloading framework for LLM-integrated vehicular networks that successfully balances model performance with privacy protection. Our approach combines federated learning with differential privacy techniques to enable secure, distributed model training across heterogeneous vehicular clients. Experimental results demonstrate that our framework achieves 75\% global accuracy while maintaining strong privacy guarantees with an optimal privacy budget of $\varepsilon = 0.8$. The minimal performance gap between private and non-private implementations (2–3\%) validates the practicality of our approach for real-world deployment. Furthermore, our framework includes a privacy-aware task partitioning algorithm and a secure communication protocol, which optimize the trade-off between local and edge computation while keeping communication overhead low and computation efficient. With computation comprising over 90\% of processing time and stable communication of approximately 2.1MB per round, the system is well-suited for resource-constrained vehicular environments. These results highlight the framework’s potential to enable safe, efficient, and privacy-preserving LLM applications in next-generation 6G vehicular networks. 

\bibliographystyle{IEEEtran}
\bibliography{refs}

\end{document}